\documentclass[10pt,aps,prd,reprint,nofootinbib,floats,floatfix,amsfonts,amssymb,amsmath,preprintnumbers,notitlepage,superscriptaddress]{revtex4-1}

\usepackage{graphicx,color} 
\usepackage{hyphenat}
\usepackage{url}
\usepackage[normalem]{ulem}

\newcommand{\sectionname}[1]{ \noindent{\bfseries #1}.---}

\begin{document}

\begin{flushleft}
KCL-PH-TH/2022-41
\end{flushleft}

\title{Dictionary learning: a novel approach to detecting  binary black holes in the presence of Galactic noise with LISA }

\author{Charles Badger}
\affiliation{Theoretical Particle Physics and Cosmology Group,  Physics Department, \\ King's College London, University of London, Strand, London WC2R 2LS, United Kingdom}
\author{Katarina Martinovic}
\affiliation{Theoretical Particle Physics and Cosmology Group,  Physics Department, \\ King's College London, University of London, Strand, London WC2R 2LS, United Kingdom}
\author{Alejandro Torres-Forn\'e}
 \affiliation{Departamento de Astronom\'ia y Astrof\'isica, 
 Universitat de Val\`encia, Dr.~Moliner 50, 46100 
 Burjassot (Val\`encia), Spain}
 \affiliation{Observatori Astron\`omic, Universitat de
 Val\`encia, Catedr\'atico Jos\'e Beltr\'an 2, 
 46980 Paterna (Val\`encia), Spain}%
\author{Mairi Sakellariadou}
\affiliation{Theoretical Particle Physics and Cosmology Group,  Physics Department, \\ King's College London, University of London, Strand, London WC2R 2LS, United Kingdom}
\author{Jos\'e A.~Font}
 \affiliation{Departamento de Astronom\'ia y Astrof\'isica, 
 Universitat de Val\`encia, Dr.~Moliner 50, 46100 
 Burjassot (Val\`encia), Spain}%
 \affiliation{Observatori Astron\`omic, Universitat de
 Val\`encia, Catedr\'atico Jos\'e Beltr\'an 2, 
 46980 Paterna (Val\`encia), Spain}%

\date{\today}

\begin{abstract}
The noise produced by the inspiral of millions of white dwarf binaries in the Milky Way may pose a threat to one of the main goals of the space-based LISA mission: the detection of massive black hole binary mergers.  We present a novel study for reconstruction of merger waveforms in the presence of Galactic confusion noise using dictionary learning. We discuss the limitations of untangling signals from binaries with total mass from $10^2 M_{\odot}$ to $10^4 M_{\odot}$. Our method proves extremely successful for binaries with total mass greater than 
$\sim 3\times 10^3$ $ M_{\odot}$ up to redshift 3 in conservative scenarios, and up to redshift 7.5 in optimistic scenarios.
In addition,  consistently good waveform reconstruction of merger events is found if the signal-to-noise ratio is approximately 5 or greater.   
\end{abstract}

\maketitle

\sectionname{Introduction}
The LIGO/Virgo interferometer network~\cite{TheLIGOScientific:2014jea,TheVirgo:2014hva} has already detected gravitational waves (GWs) from almost one hundred compact binary coalescence (CBC) events~\cite{abbott2021gwtc1,abbott2021gwtc2,abbott2021gwtc3}. These detections populate GW transient catalogues and reveal information about properties of the underlying black hole and neutron star populations~\cite{LIGOScientific:2020kqk}. The high frequency range probed by the current terrestrial detectors, at around $(10-1000)$ Hz, is sensitive to stellar-mass binaries that mostly lie below the pair-instability supernova mass gap\footnote{GW190521 is the only exceptional event where the mass of the primary black hole is unambiguously in the mass gap~\cite{GW190521D}.}. Mergers of supermassive black holes, however, are expected to emit low-frequency (mHz) gravitational waves. 

The space-based GW interferometer LISA, anticipated to be launched in the mid-2030s, will be sensitive to GWs in the mHz range~\cite{LISA}. Other GW sources will be detectable at these frequencies: Galactic white dwarf binaries, inspiraling binaries with extreme mass-ratio, or colliding true vacuum bubbles formed at the electroweak phase transition~\cite{Amaro-Seoane:2012aqc, Robson:2018ifk, Caprini:2015zlo}. 
The tens of millions of double white dwarf binaries in the Galaxy could have an impact on detectability of massive black hole binaries coalescing in the LISA frequency band~\cite{Ruiter:2007xx}. LISA will observe continuous GWs from inspiraling white dwarfs, and although it may be sensitive to individual sources, most will remain unresolved and these are referred to as Galactic confusion noise~\cite{Timpano:2005gm, LISA_Big, Nissanke_2012}. It has been shown that modulation of the Galactic foreground from the LISA orbit could lead to a reduction in signal-to-noise ratio (SNR) of other GW sources by a factor of 4~\cite{Seto:2004ji}. A LISA Data Challenge\footnote{\url{https://lisa-ldc.lal.in2p3.fr}.} is underway to study the impact of overlapping Galactic sources on the sensitivity to massive black hole mergers~\cite{LSD}, and attempts to separate the foreground from other GW sources have been conducted~\cite{Adams_2014, Boileau_2021, Barack_2004, Boileau_2022, Pieroni:2020rob}.

In this {\sl Letter} we apply a dictionary learning method to separate CBCs from the Galactic foreground in the LISA frequency band. Such a method has been successfully applied in GW data analysis to classify and denoise Advanced LIGO's ``blip'' noise transients~\cite{Torres-Forne:2020eax} and effectively improve the performance of the detector. More precisely, we assess the suitability of the dictionary learning method for the classification and reconstruction of massive binary black hole merger signals in the presence of Galactic noise. 

Previous studies focused on the inspiral of loud CBC sources 
and demonstrated that SNR accumulated over time is sufficiently large to overcome the noise from Galactic binaries~\cite{Cornish:2006ms}. In other literature, detectability of CBCs was investigated for equal-mass and non-spinning binaries~\cite{Klein:2015hvg,Sesana:2021jfh,LISA:2017pwj}, confirming the largest SNR is expected from binaries with combined mass $\sim (10^5-10^6) M_{\odot}$. In particular, Fig.~3 in~\cite{LISA:2017pwj} presents two mass ranges with low SNR that could be affected by Galactic foreground, namely $(10^2-10^4) M_{\odot}$ and $(10^7-10^9) M_{\odot}$. 

Here we consider all of the mass ranges, along with varying spins and redshifts, and we study their waveforms around coalescence time. The dictionary learning method reconstructs CBC signals with ease in the trivial case where the CBCs are above the Galactic noise, i.e. for the $(10^5-10^6) M_{\odot}$ mass range. We find the dictionary learning method to be too computationally expensive for very heavy mergers in the  range $(10^7-10^9) M_{\odot}$. However, our method succeeds in separating low-SNR binaries in the range $(10^2-10^4) M_{\odot}$ from the Galactic noise. Hence, the dictionary learning method could significantly assist the detection of this prime LISA source~\cite{Barausse:2020mdt}.

\sectionname{Dictionary learning}
Any CBC signal in the LISA band will be overlaid with continuous waves from the inspiral of double white dwarfs. Therefore, we can model the detector strain, $y(t)$, as a superposition of the CBC signal $u(t)$ and the Galactic confusion noise $n(t)$:
\begin{equation}
    y(t)=u(t)+n(t).
\end{equation}
We express the loss function as
\begin{equation}
\label{eq:constrain}
    J(u) = ||y-u||_{L_2}^2 + \lambda \mathcal{R}(u),
\end{equation}
and search for a solution $u_{\lambda}$ that minimises $J(u)$, where $||\cdot||_{L_2}$ is the $L_2$ norm~\cite{10.1137/07070156X,10.1145/1553374.1553463}.
The first term in the loss function, often referred to as the {\sl error term}, measures how well the solution fits the data, while the regularisation term $\mathcal{R}(u)$ captures any imposed constraints. 
The regularisation parameter 
$\lambda$ tunes the weight of the regularisation term relative to the error term; it is a hyperparameter of the optimisation process. 

The goal of the dictionary learning method~\cite{dict_learning} is to find the sparse vector $\alpha$ that reconstructs the true signal $u$ as a linear combination of columns of a dictionary $\textbf{D}$,
\begin{equation}
u \sim \textbf{D} \alpha,
\end{equation}
with $\textbf{D} $ a matrix of prototype signals (atoms) trained to reconstruct a given set of signals, which for our study is CBCs. Sparsity of the vector $\alpha$ is imposed via the regularisation term $\mathcal{R}(u)=||\alpha||_{L_1}$, using the $L_1$ norm. Therefore, the constrained variational problem in (\ref{eq:constrain}) reads
\begin{equation}
\label{eq:LASSO}
    \alpha_{\lambda} = \underset{\alpha}{\rm{argmin}}\left\{||y- \textbf{D}\alpha||_{L_2}^2 + \lambda ||\alpha||_{L_1}\right\},
\end{equation}
and is called ``basis  pursuit''~\cite{basis_pursuit}  or ``least absolute shrinkage and selection operator''
(LASSO)~\cite{lasso_paper}. 

The basis pursuit can be improved significantly if, instead of using a predefined dictionary, we apply a learning process where the dictionary is trained to fit a given set of signals. The procedure starts by selecting templates of CBC waveforms and whitening the data. The waveforms are aligned at the strain maximum and divided into patches, with the number of patches ($p$) much larger than the length of each patch ($d$). To train the dictionary we  solve  (\ref{eq:LASSO}) considering both the sparse vector $\alpha$ and the dictionary $\textbf{D}$ as variables:
\begin{equation}
\label{eq:dict_learning}
\alpha_{\lambda}, \textbf{D}_{\lambda}=\underset{\alpha, \textbf{D}}{\rm{argmin}} \left\{\frac{1}{d}\sum_{i=1}^{p}||\textbf{D}\alpha_i- {x}_i||^2_{L_2}+\lambda ||\alpha_i||_{L_1}\right\},
\end{equation}
with $x_i$ denoting the $i$-th training patch. This problem is not jointly convex unless the variables are considered separately as outlined in~\cite{Mairal:2009}.

In our study we create training signals that contain CBC waveforms only and no noise. The dictionary created is then tested on signals that include new CBC waveforms \textit{combined} with Galactic noise. We describe briefly the massive black hole and white dwarf binary waveforms used in our datasets below.

\sectionname{Training and testing datasets}
We utilise the \texttt{IMRPhenomD} approximant~\cite{Khan:2015jqa} provided by the LISA Data Challenge to model waveforms of binary black holes detectable by LISA, capturing inspiral, merger and ringdown of the signal. Binaries with total mass $(10^5-10^6) M_{\odot}$ are expected to have $\rm{SNR} \geq 150$, making separation from the Galactic foreground a trivial problem. CBCs in the mass range $(10^7-10^9) M_{\odot}$ have lower frequencies, making it difficult for the dictionary learning to reconstruct their sinusoidal behavior. We thus study reconstruction capabilities of binary black holes with total mass ranging from $(10^2-10^4) M_{\odot}$. The dictionary is trained on a set of 100 noiseless CBC signals, simulated over one day with cadence\footnote{The cadence was chosen low enough to have a high sampling rate that avoids aliasing, but high enough to allow for reasonable computational time.} $\Delta t = 2$ s. 
Table~\ref{table:cbc_params} lists the relevant parameters of the \texttt{IMRPhenomD} waveform and the corresponding ranges of values we choose for the CBC sources. We simulate the data by drawing randomly from the probability distribution of the parameters. The redshift for all sources is fixed to $z=2$, since changing the redshift leads to a simple rescaling of the amplitude that has no impact on our whitened data in the training set. Note that the same does not hold for the testing data, since changing redshift would change the relative amplitude of the CBCs to the Galactic foreground.  
\begin{table}
	\begin{tabular}{ll}
	   Parameter &  Distribution \\ \hline
		Total mass ($M_\odot$)  &  logU[$10^2,10^4]$ \\[4pt]
		Mass ratio   &   U[1,10] \\[4pt]
		Primary spin  &  U[-1,1] \\[4pt] 
		Secondary spin  &    U[-1,1]    \\[4pt]
		Redshift &  2 \\[4pt]
		Luminosity Distance (Mpc) &  15975 \\[4pt] 
	\end{tabular}
	\caption{Parameters used to construct training CBC signals with the \texttt{IMRPhenomD} waveform approximant. We choose values randomly from the uniform distributions indicated in the right column, keeping redshift and luminosity distance fixed.}
	
	\label{table:cbc_params}
\end{table}

Consider two white dwarfs of mass $M_1$ and $M_2$ on a quasi-circular orbit with inclination $\iota$ at a distance $R$. They emit GWs with amplitude ~\cite{DWD_Eq}
\begin{multline}
    A_+(\mathcal{M}_c, R, f_{\rm{GW}}, \iota) = \frac{2G^{5/3}\mathcal{M}_c^{5/3}}{c^4R} (\pi f_{\rm{GW}})^{2/3} (1+\cos^2\iota), \\
    A_\times(\mathcal{M}_c, R, f_{\rm{GW}}, \iota) = -\frac{4G^{5/3}\mathcal{M}_c^{5/3}}{c^4R} (\pi f_{\rm{GW}})^{2/3} \cos\iota,
    \label{eq: DWD_Amp}
\end{multline}
for the $+$ and $\times$ polarisations, respectively. The chirp mass of the binary is a function of the progenitors' masses, $\mathcal{M}_c = (M_1M_2)^{3/5}/(M_1+M_2)^{1/5}$, and the GW frequency is twice the orbital frequency, $f_{\rm{GW}} = 2f_{\rm{orb}}$. 
The resulting plane wave has a slight frequency shift over time, and for each polarisation reads
\begin{eqnarray}
    h_+(t) &=& A_+ \cos(2\pi f_{\rm{GW}}t + \dot{f}_{\rm{GW}}t^2 + \phi_0)\,, \\
    h_{\times}(t) &=& A_{\times} \sin(2\pi f_{\rm{GW}}t + \dot{f}_{\rm{GW}}t^2 + \phi_0)\,,
    \label{eq: DWD_waveform}
\end{eqnarray}
where $\phi_0$ stands for the initial phase and $\dot{f}_{\rm{GW}}$
 for the GW frequency time derivative
\begin{equation}
    \dot{f}_{\rm{GW}} = \frac{96}{5}\bigg(\frac{G\mathcal{M}_c}{c^3}\bigg)^{5/3} \pi^{8/3} f^{11/3}_{\rm{GW}}.
    \label{eq: DWD_freqDeriv}
\end{equation}
To simulate the LISA Galactic foreground we sum over GW signals from the  white dwarf binaries in our galaxy:
\begin{equation}
    n(t) = \sum_{i=1}^N \sum_{A=+,\times} \frac{\sqrt{3}}{2}F_A h_{A,i}(t),
    \label{eq: DWD_totalSig}
\end{equation}
where $F_{+,\times}$ stands for the detector response function~\cite{Cutler_1998}. 
The mass, location, and orbital frequency of nearly 5 million
white dwarf binaries in the Milky Way are taken from  \cite{Lamberts:2019nyk}. 

\begin{figure}
    \centering
    \includegraphics[width=\linewidth]{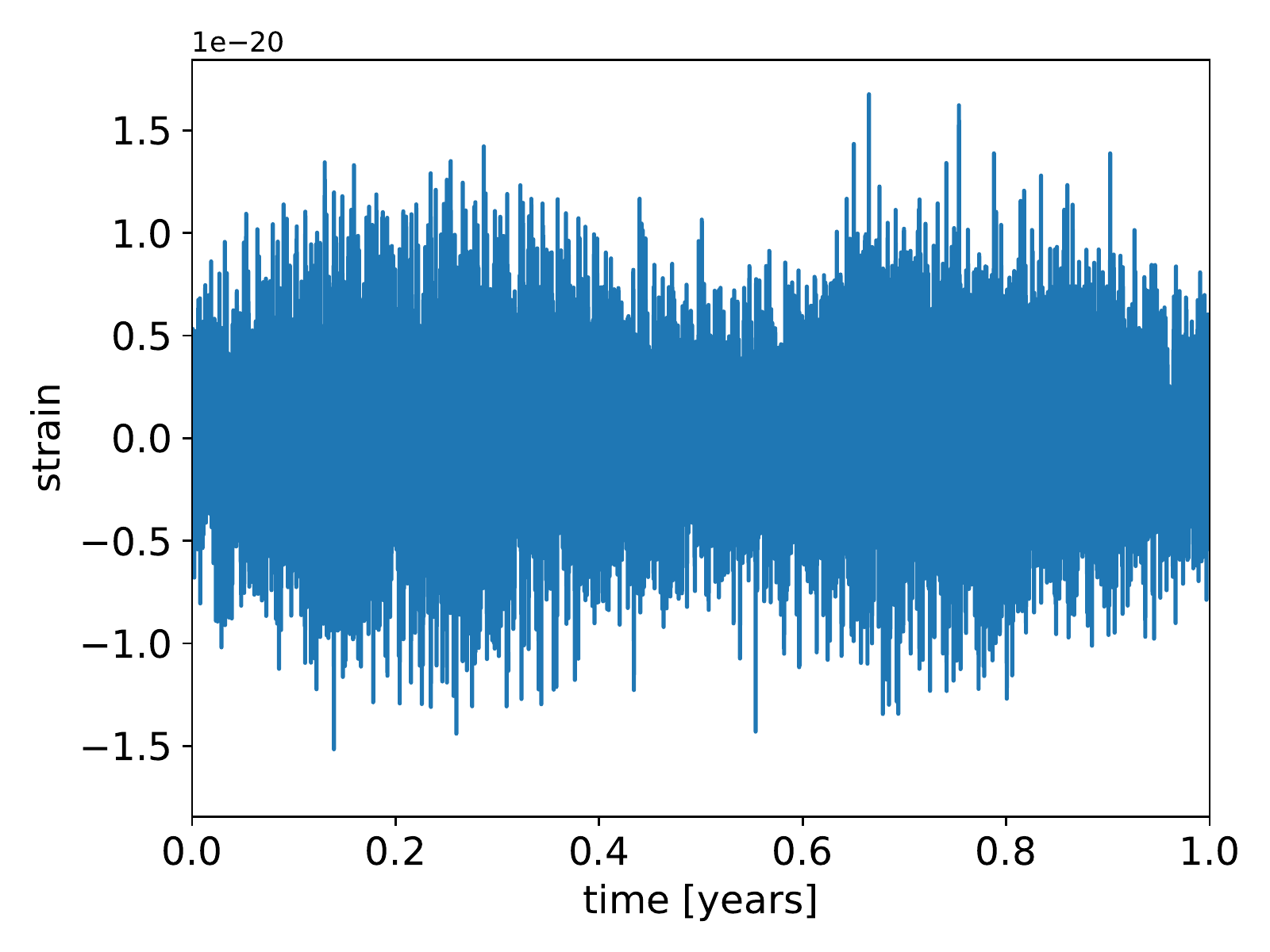}
    \caption{Strain due to the inspiral of binary white dwarfs in the Galaxy as measured by LISA over a 1 year orbit. The modulation from LISA's orbit creates peaks at $1/4$ and $3/4$ of the year, when the normal of LISA's constellation plane is pointed at or away from the Galactic center.}
    \label{fig:DWD_back}
\end{figure}
We present the resulting Galactic foreground in Fig.~\ref{fig:DWD_back}. The orbit of LISA around the Sun introduces a modulation in the Galactic noise (\ref{eq: DWD_totalSig}), with maximum value when the normal of the LISA constellation plane is closest to the binary location, as expected since white dwarfs are supposed to cluster near the Galactic center. In our analysis, we will first consider the Galactic foreground at a time of the year when it is maximum. We subsequently consider testing signals that combine Galactic foreground and a CBC signal.


\sectionname{Results}
The hyperparameters of dictionary learning, namely the regularisation parameter $\lambda$, the atom (and patch) length $d$ and the number of patches $p$, have an impact on the quality of the signal reconstruction. We fix $p=3 \, d/2$ to ensure a complete dictionary (one where the number of atoms is greater than atom length) and choose $d \in [2^2,2^7]$. Independently of our choice of $d$, we find the optimal regularisation parameter to lie in the range $\lambda_{\rm opt} \in [10^{-3}, 10^{-2}]$ (see Supplemental Material for a detailed study). 
For the remainder of the analysis we fix $\lambda=10^{-3}$, as little quantitative differences to our results are found with the choice of $\lambda=10^{-2}$. 

To find the best dictionary size, for each reconstruction we calculate the so-called overlap between the injected CBC waveform $h_i(f)$ and the recovered waveform $h_r(f)$,
\begin{equation}
    \mathcal{O} = \frac{(h_i|h_r)}{\sqrt{(h_i|h_i)(h_r|h_r)}},
\end{equation}
with
\begin{equation}
    (x|y)=2 \int^\infty_0 \frac{x(f) y^*(f) + x^*(f) y(f)}{S_n(f)} {\rm{d}}f,
\end{equation}
where $S_n(f)$ is the one-sided noise power spectral density. The overlap $\mathcal{O}$ can range between -1 and 1, with 1 reflecting perfectly matched signals, and -1 implying perfect anti-correlation. The overlap is widely used in the  GW community for identifying transient CBC signals through matched filtering using waveform template banks \cite{Cutler:1994ys,Cornish:2014kda, LIGOScientific:2019hgc, Cornish:2020dwh}.  
We track this metric across the testing dataset and choose the optimal atom length $d_{\rm opt}$ that maximises it, namely $d_{\rm opt}=4$
(see Supplemental Material).

To have a metric for error, we also calculate the overlap between the recovered CBC waveform and the present Galactic foreground $n$, which we denote as $\mathcal{O}(h_r, n)$. In addition, we define the overlap difference $\Delta \mathcal{O} = \mathcal{O}(h_r, h_i) - \mathcal{O}(h_r, n)$ to evaluate how much more the reconstructed signal has in common with the injected signal than with Galactic foreground.

\begin{figure}
    \centering
    \includegraphics[width=\linewidth]{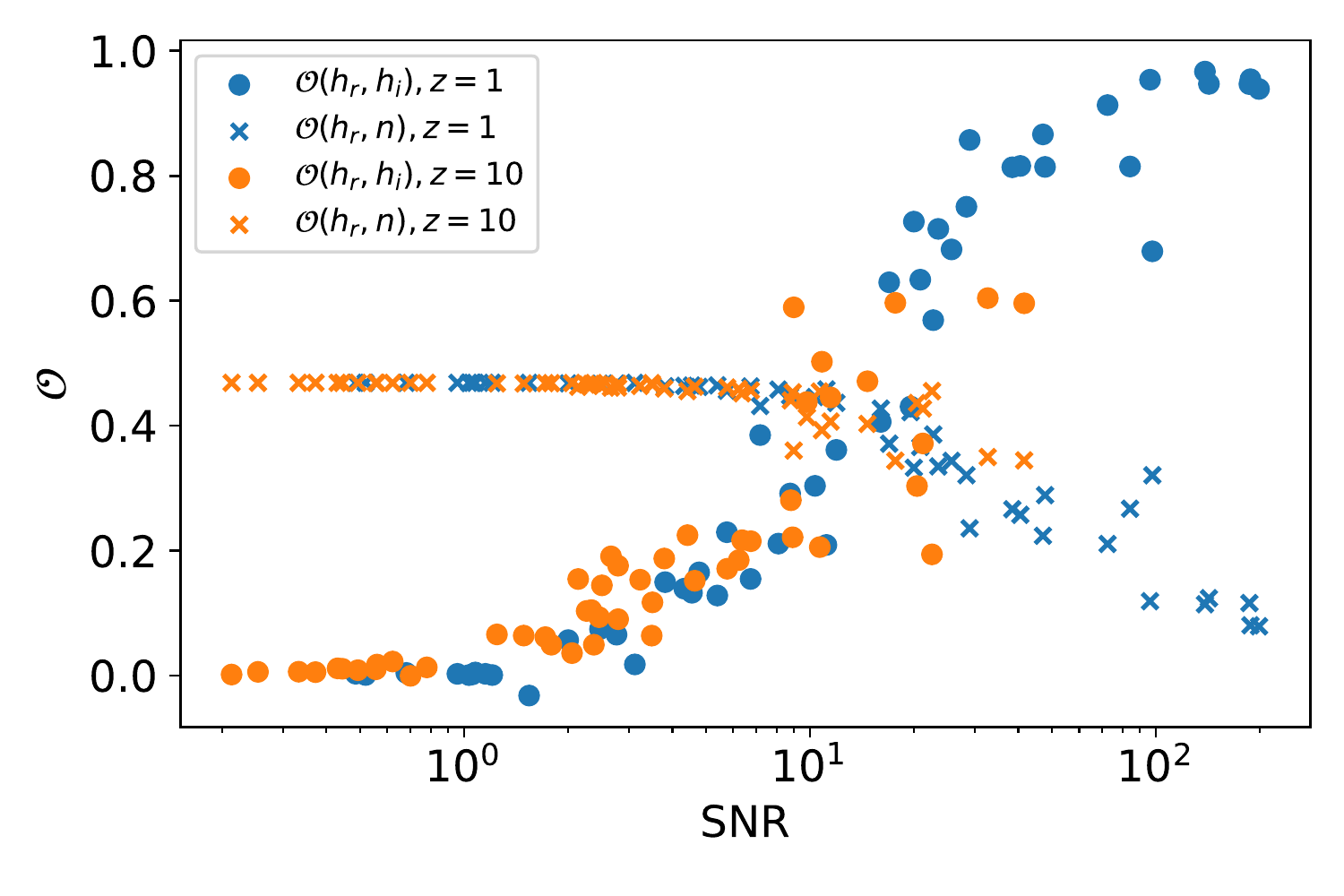}
    \caption{Overlap between reconstructed signal and injected CBC waveform as a function of SNR for redshift $z=1,10$ and fixed atom length $d=4$, overlayed with recovery overlap between reconstructed signal and noise for the same data.}
    \label{fig:OSig_Max}
\end{figure}

We now study how well the CBCs can be reconstructed using a dictionary with $d=4$ and $\lambda = 10^{-3}$. Specifically, we create two sets of 50 signals with CBCs at redshift $z=1$ and $z=10$, and investigate how their reconstruction varies with SNR. From Fig.~\ref{fig:OSig_Max} we see that the overlap between reconstructed signal and injected CBC waveform (solid circles) increases with SNR, while the overlap between reconstructed signal and noise (crosses) decreases with SNR, as expected. Interestingly, the noise overlap is approximately constant until SNR $\approx 10$, where it starts to decrease, while the reconstruction starts to improve significantly, with $\mathcal{O}(h_r,h_i)= 1$ for some of the $z=1$ waveforms. Note that the $z=1$ and $z=10$ datasets do not differ greatly in overlap, and when they do it is between sources of very different mass. Therefore, we turn to study how overlap changes as a function of both redshift and total mass of the CBC.

\begin{figure}
    \centering
    \includegraphics[width=\linewidth]{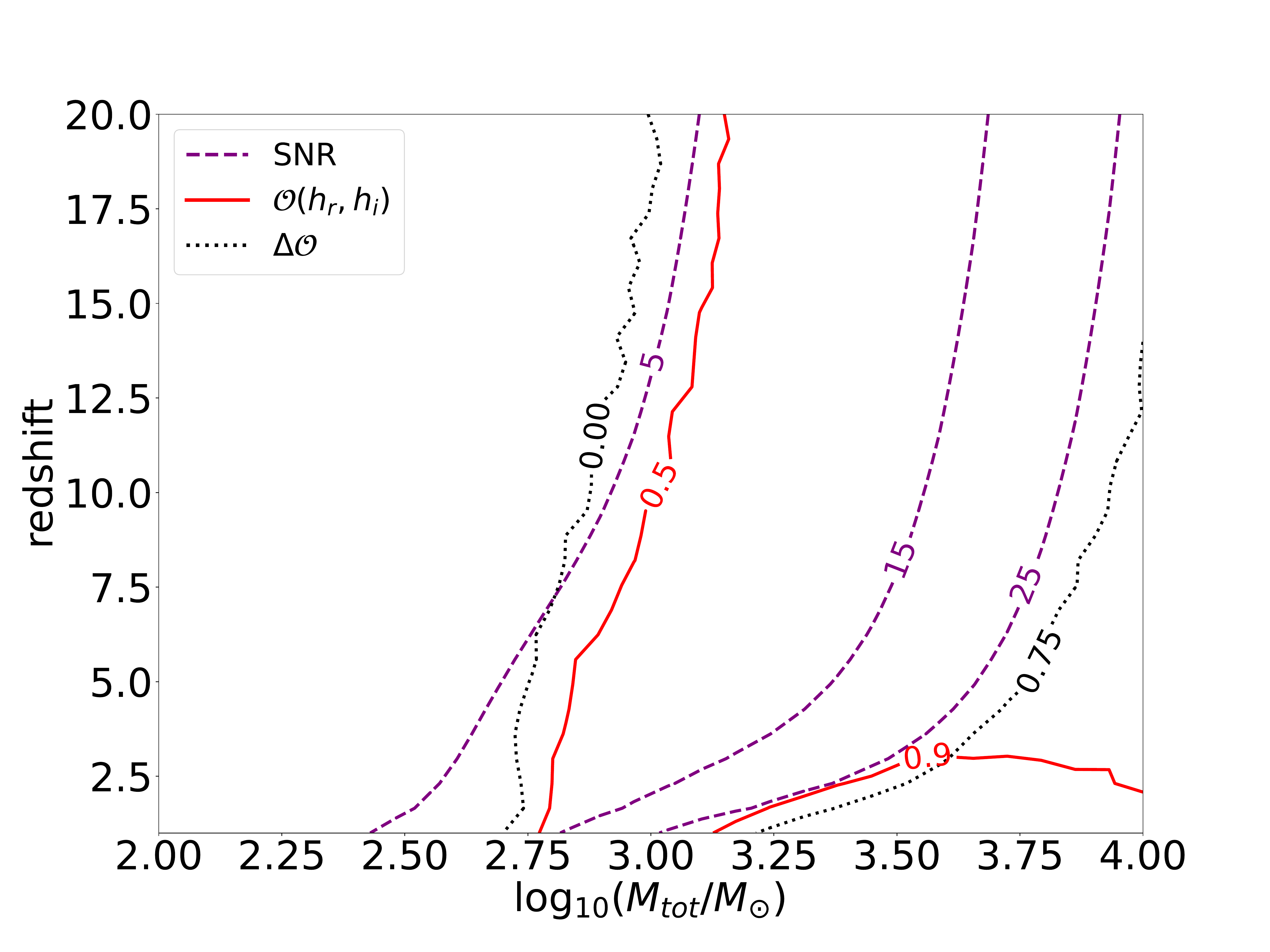}
    \includegraphics[width=\linewidth]{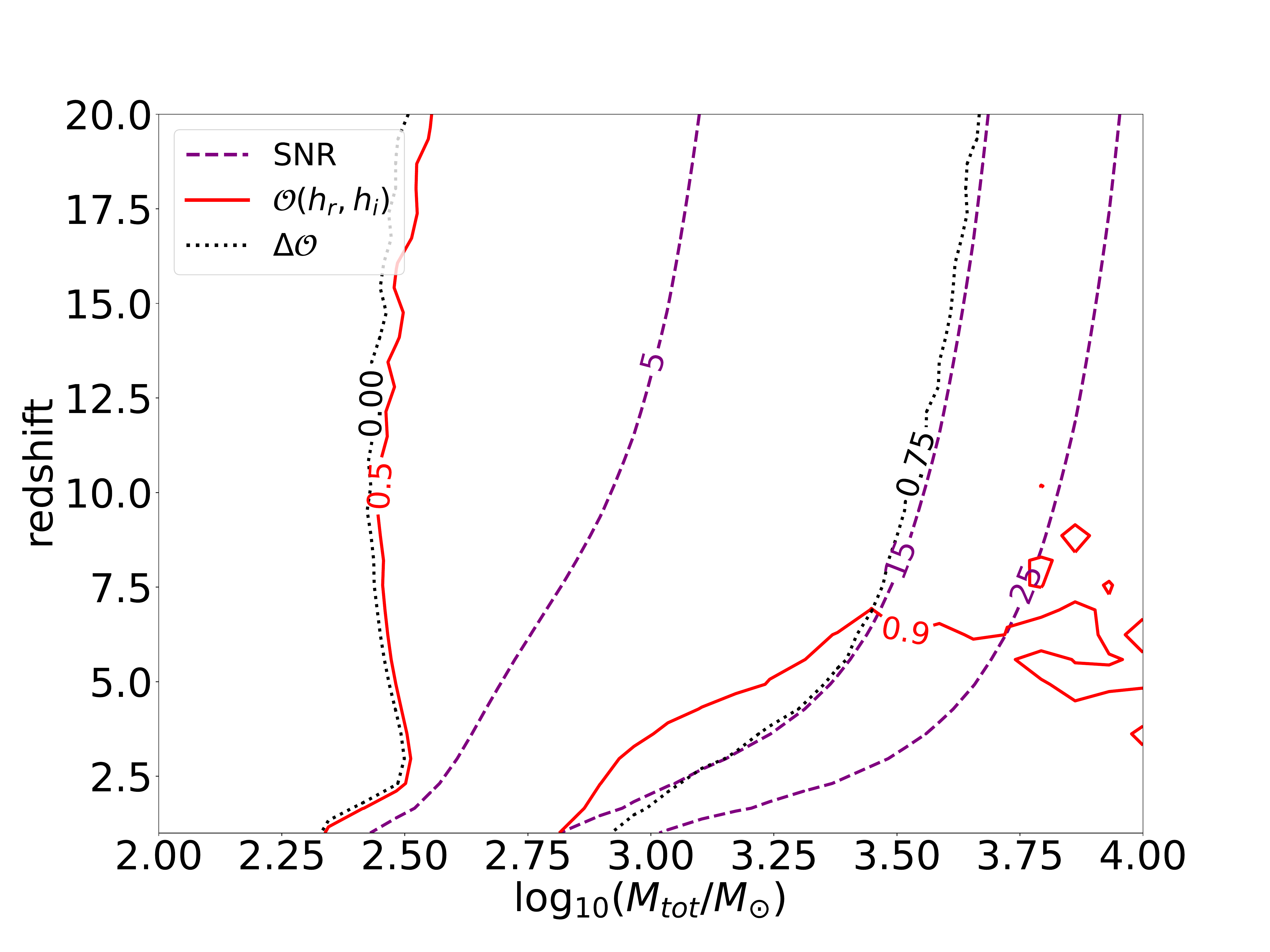}
    \caption{Overlap between reconstructed signal and injected CBC waveform as a function of total mass and redshift in strong (top) and weak (bottom) Galactic foreground scenarios, fixing binary mass ratio to 1 and black hole spins to 1.}
    \label{fig:Mvz_MinMax}
\end{figure}

We begin by fixing mass ratio to 1 and both black holes' spins to 1, for simplicity, and we create a dataset of 400 CBC events with uniform spacing $1\leq z \leq 20$ and log-uniform spacing for total mass of the binary $10^2 M_{\odot} \leq M_{\rm tot} \leq 10^4 M_{\odot}$. Each event is overlayed 
with Galactic foreground at a yearly modulation maximum and minimum (see Fig.~\ref{fig:DWD_back}), and reconstructed. Resulting contours of signal overlap $\mathcal{O}(h_r, h_i) = 0.5$, $0.9$, overlap difference $\Delta\mathcal{O} = 0$, $0.75$ and $\rm{SNR} = 5$, $15$, $25$ are plotted in Fig.~\ref{fig:Mvz_MinMax}. Although generally speaking increasing SNR improves reconstruction capabilities, reconstruction success is more dependent on the CBC's total mass and redshift. 

In strong (maximum) Galactic noise scenarios, binaries with total mass greater than $1330$ $M_{\odot}$ can be reconstructed with $\mathcal{O}(h_r, h_i) > 0.5$ for the redshift range. Reconstructions with $\mathcal{O}(h_r, h_i) > 0.5$ in weak (minimum) Galactic foreground scenarios can be achieved for binaries with total mass greater than $355$ $M_{\odot}$.
Extremely good signal reconstruction, with overlap greater than 0.9, can be achieved for sources with total mass greater than 1350 $M_{\odot}$ up to redshifts of 3 in the pessimistic case, and up to redshifts as large as 7.5 in the optimistic case. 
From Fig.~\ref{fig:Mvz_MinMax} one can also track how the overlap difference $\Delta\mathcal{O}$ varies with binary mass and redshift. All sources that lie in the parameter space at the right of the  $\Delta\mathcal{O}=0$ contour lead to reconstructed signals that are more similar to the true, injected signal than the Galactic noise. In the pessimistic case this is true for total masses greater than 1000 $M_{\odot}$, and in the optimistic case for total masses as small as $315$ $M_{\odot}$.

\sectionname{Conclusions}
Gravitational-wave signals from the inspiral of white dwarf binaries in the Galaxy must be considered when studying detection of CBC signals with LISA~\cite{Ruiter:2007xx}. In this work we have modelled this Galactic foreground, or confusion noise, and studied its impact on the reconstruction of massive black hole binary merger signals using an approach based on learned dictionaries. We have found dictionary learning to be a promising technique for detection of such signals in the presence of Galactic foreground. Reconstructing the CBC waveforms with dictionary learning can be optimised with atom length $d=4$ and regularisation parameter $\lambda \in [10^{-3}, 10^{-2}]$. 

The threshold overlap between the injected and  the reconstructed signal, typically chosen to be $\mathcal{O}(h_r, h_i)=0.5$, can be achieved for binaries with mass $M_{\rm tot} > 1330 M_{\odot}$ in strong Galactic foreground, and with mass $M_{\rm tot} > 355 M_{\odot}$ in weak Galactic foreground at all redshifts. For CBCs with total mass $M_{\rm tot} > 3150 M_{\odot}$, the reconstructed waveform and the true waveform overlap greatly with $\mathcal{O}(h_r, h_i)>0.9$ up to redshift $z=3$ and $z=7.5$ in the case of strong and weak Galactic foreground, respectively. For all tested signals, we calculate the overlap between the reconstructed CBC waveform and the Galactic noise, and its difference to the overlap between the reconstructed and injected CBC waveform. We conclude that the reconstructed signal overlaps more with the true CBC signal than with the noise for binaries with $M_{\rm tot} > 1000 M_{\odot}$.

Further assessment of the dictionary-learning approach presented in this {\it Letter} to assist detection of other types of GW sources in the LISA frequency band 
will be reported elsewhere. Those investigations include the analysis of extreme mass ratio inspirals and the case of overlapping CBC sources.

\acknowledgments{The authors are grateful to Astrid Lamberts for providing a catalogue of double white dwarf sources that allowed us to model the Galactic background. The authors also thank Nelson Christensen for providing helpful feedback.
A.T.-F., M.S. and J.A.F. thank the Institute for Pure \& Applied Mathematics (IPAM), University of California Los Angeles (UCLA) where this project was initiated at the occasion of the {\sl Mathematical and Computational Challenges in the Era of Gravitational-Wave Astronomy} workshop.

We acknowledge computational resources provided by the LISA Data Challenge working group in the LISA consortium. The software packages used in this study are \texttt{matplotlib}~\cite{Hunter:2007}, \texttt{numpy}~\cite{numpy}, and \texttt{MATLAB Signal Processing Toolbox}~\cite{matlabsignal}.

K.M. is supported by King's College London through a Postgraduate International
Scholarship. M.S. is supported in part by the Science and Technology Facility Council (STFC), United Kingdom, under the research grant ST/P000258/1. A.T.-F. and J.A.F. acknowledge support from the Spanish Agencia Estatal de  Investigaci\'on  (PGC2018-095984-B-I00 and PID2021-125485NB-C21) and  by  the Generalitat Valenciana (PROMETEO/2019/071).

}

\bibliography{ml}


\end{document}


\title{Dictionary learning: a novel approach to detecting  binary black holes in the presence of Galactic noise with LISA - Supplemental Material}

\author{Charles Badger}
\affiliation{Theoretical Particle Physics and Cosmology Group,  Physics Department, \\ King's College London, University of London, Strand, London WC2R 2LS, United Kingdom}
\author{Katarina Martinovic}
\affiliation{Theoretical Particle Physics and Cosmology Group,  Physics Department, \\ King's College London, University of London, Strand, London WC2R 2LS, United Kingdom}
\author{Alejandro Torres-Forn\'e}
 \affiliation{Departamento de Astronom\'ia y Astrof\'isica, 
 Universitat de Val\`encia, Dr.~Moliner 50, 46100 
 Burjassot (Val\`encia), Spain}
\author{Mairi Sakellariadou}
\affiliation{Theoretical Particle Physics and Cosmology Group,  Physics Department, \\ King's College London, University of London, Strand, London WC2R 2LS, United Kingdom}
\author{Jos\'e A.~Font}
 \affiliation{Departamento de Astronom\'ia y Astrof\'isica, 
 Universitat de Val\`encia, Dr.~Moliner 50, 46100 
 Burjassot (Val\`encia), Spain}%
 \affiliation{Observatori Astron\`omic, Universitat de
 Val\`encia, Catedr\'atico Jos\'e Beltr\'an 2, 
 46980 Paterna (Val\`encia), Spain}%

\date{\today}

\maketitle

\begin{figure}
    \centering
    \includegraphics[width=\linewidth]{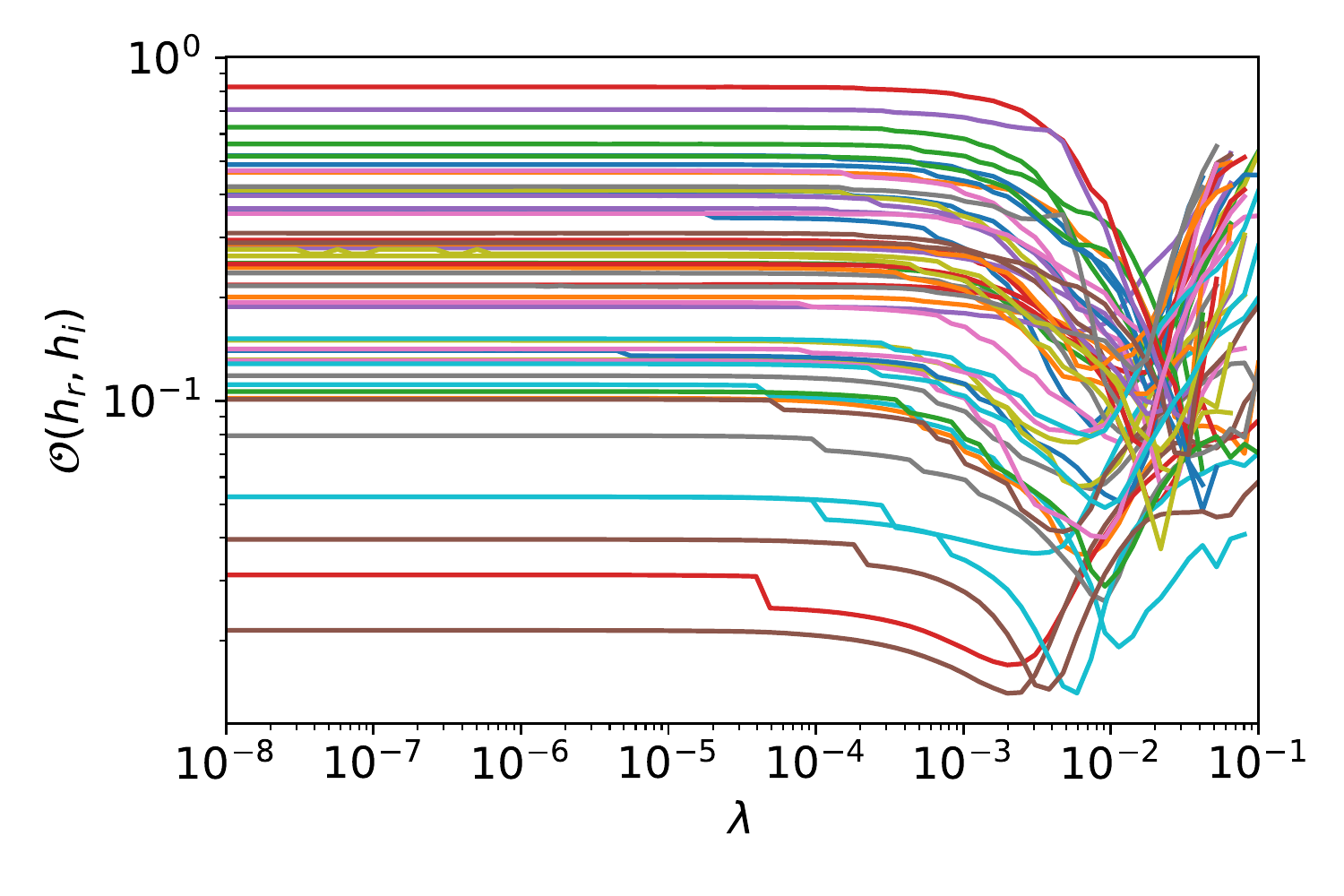}
    \caption{Optimal value of the regularisation parameter is between $10^{-3}$ and $10^{-2}$. For $\lambda < 10^{-3}$ the $\mathcal{O}$ curve plateaus.}
    \label{fig:d64}
\end{figure}

\sectionname{Optimal regularisation parameter}
In theory there exists an optimal regularisation parameter $\lambda_{\rm opt}$ such that we retrieve the best reconstruction of a CBC signal. For each signal in the testing dataset, we vary $\lambda \in [10^{-8}, 10^{-1}]$ and reconstruct the CBC waveform. The calculated overlap between the reconstructed signal and the injected CBC,  $\mathcal{O}$, can be seen in Fig.~\ref{fig:d64}. For all reconstructions with $\lambda \lesssim 10^{-3}$, $\mathcal{O}$ plateaus. Some cases ``dipped" to a minimum $\mathcal{O}$ before dramatically increasing as $\lambda$ increased. Regardless of the $\mathcal{O}$ behavior each CBC demonstrated, the optimal regularisation parameter $\lambda_{\rm opt}$ took on a value between $10^{-3}$ and $10^{-2}$. This behavior was consistent over differing atom length $d$ and Galactic noise strength.

\FloatBarrier

\sectionname{Optimal atom length}
The length of the columns of the dictionary significantly impacts quality of reconstruction. 
We fix $\lambda=10^{-3}$ and vary dictionary size $d \in [2^2,2^7]$ to find the value of atom length that results in best reconstruction of the CBC signals. In Table \ref{tab:d_opt} we report the change in the sum of $\mathcal{O}$s for all 50 individual testing samples as we alter the atom length. With this we identify the optimal atom length to be $d_{\rm opt}=4$.

\FloatBarrier
\begin{table}
    \centering
    \begin{tabular}{c|c}
       atom length, $d$  &  $\sum_i \, (\mathcal{O}_i)$ \\
       \hline
        4 & 14.3 \\
        8 & 10.2 \\
        16 & 6.65 \\
        32 & 5.65 \\
        64 & 6.95. \\
        128 & 7.93
    \end{tabular}
    \caption{Variation in the total sum of overlaps as a function of atom length. The optimal atom length that gives best reconstruction is $d_{\rm opt}=4$.}
    \label{tab:d_opt}
\end{table}